\documentclass[aps,pra,twocolumn,showpacs,epsfig]{revtex4-1}

\usepackage{graphicx}
\usepackage{amsmath}
\usepackage{epsfig}
\usepackage{dcolumn}
\usepackage{bm}
\usepackage{color}
\usepackage{epstopdf}

\newcommand{\ket}[1]{\left\vert#1\right\rangle}
\newcommand{\bra}[1]{\left \langle#1\right\vert}

\begin{document}


\title{Non-Markovianity, Loschmidt echo and criticality: a unified picture}

\author{P.~Haikka$^{1}$, J.~Goold$^{2,3}$, S.~McEndoo$^{1,4}$, F.~Plastina$^{5,6}$, S.~Maniscalco$^{1,4}$}

\affiliation{$^1$ Turku Center for Quantum Physics, Department of
Physics and Astronomy, University of Turku, FIN20014, Turku,
Finland} 
\affiliation{$^2$Clarendon Laboratory, University of
Oxford, Oxford, United Kingdom,} 
\affiliation{$^3$Physics Department,
University College Cork, Cork, Ireland,}
 \affiliation{$^4$ SUPA,
EPS/Physics, Heriot-Watt University, Edinburgh, EH14 4AS, United Kingdom,}
\affiliation{$^5$ Dipartimento di Fisica, Universit\`{a} della
Calabria, 87036 Arcavacata di Rende (CS), Italy,}
\affiliation{$^6$ INFN--Gruppo Collegato di Cosenza, Italy}

\date{\today}

\begin{abstract}
A simple relationship between recently proposed measures of
non-Markovianity and the Loschmidt echo is established, holding
for a two-level system (qubit) undergoing pure dephasing due to a
coupling with a many-body environment. We show that the Loschmidt echo is intimately related to the
information flowing out from and occasionally back into the system. This, in turn, determines the non-Markovianity of the reduced dynamics. In particular, we consider a central qubit
coupled to a quantum Ising ring in the transverse field. In this context, the information
flux between system and environment is strongly affected by the environmental criticality; the qubit dynamics is shown to be Markovian exactly and only at the critical point. Therefore non-Markovianity is an indicator of criticality in the model considered here.
\end{abstract}
\maketitle

The process of decoherence is not only of paramount importance to
understand the emergence of the classical world from the
microscopic domain of quantum mechanics, but it also constitutes
the fundamental obstacle for the design of reliable quantum
technologies \cite{Joos,Zurekrmp,Schlosshauer}. By now,
decoherence is understood to be the result of a relentless
monitoring of the quantum system by an environment. In this
description a preferred set of states, known as the pointer basis,
is singled out in a process known as environment induced
superselection or einselection~\cite{zurek:82,zurek2:82}.
Essentially, interactions between the system and the environment
lead to an entangled composite state and phase relations between
the pointer states are lost. In other words, the system is giving
up information which otherwise would inhibit the emergence of
classicality.

The dynamics of the reduced system state is typically described
using the tools of open quantum systems theory~\cite{bandp}. A
process where the information flows out continuously from the
system (and is, in fact, delocalized in the correlations) is known
as a quantum Markov process. In such a case, the reduced state
dynamics may be described using a semi-group of completely
positive dynamical maps, equivalent to a quantum master equation
in Lindblad form. The microscopic derivation of the master
equation often requires severe approximations, such as weak
interaction between system and environment, and crucially the
assumption of a fast decaying temporal correlation function of the
environment \cite{Linblad}. This approach has been remarkably
successful to describe a wealth of processes in the field of
quantum optics. However, when strong-coupling and structured or
finite size environments are concerned, one often encounters
non-Markovian behavior, with a dynamics characterized by memory
effects whereby during certain time intervals a reflux of
information back to the system is observed.

The characterization of a given quantum dynamical process in terms
of its non-Markovianity has recently been the subject of intense
theoretical work resulting in the construction of a number of
quantitative measures \cite{wolf:08,elsi:09,Plenio:10,Fisher:10}.
In particular, the measure by Breuer \textit{et al.}
\cite{elsi:09}, based on the trace distance, has been used to
experimentally explore non-Markovianity in a linear optics
setup \cite{Lieu}, and to investigate non-Markovianity in a variety
of different system-environment models which include spin-chains
\cite{maurospin,maurospin2}, biomolecules \cite{biology} and
ultra-cold quantum gases \cite{pinjacold}.

Interacting many-body environments are highly structured and
typically induce non-Markovian dynamics on systems interacting
with them. Therefore it can be difficult to obtain a full
description of the reduced dynamics and of the information flux.
Where pure dephasing of a qubit is concerned, one can express the
reduced qubit state in terms of the Loschmidt echo
\cite{zurek:03}, which gives a measure of the environmental
response to the perturbation induced by system \cite{Peres,Gorin}.
The Loschmidt echo has been extensively used to explore
decoherence dynamics for critical environments
\cite{Quan,fernando,fazio,cecilia1,cecilia2,Zurek11,Goold2011}
where it has been shown that in the weak coupling regime, the
decoherence rate is greatly enhanced.

Our aim is twofold: {\it i}) we will study the general
relationships between the Loschmidt echo, the measures of
non-Markovianity and the master equation for a qubit undergoing
pure dephasing; and {\it ii}) we will illustrate these concepts by
revisiting, from the viewpoint of information flux, the
hypersensitivity to environmental criticality of the decoherent
dynamics of a qubit centrally coupled to a quantum spin chain in a
transverse field \cite{Quan}. We will express the information flow
in terms of the Loschmidt echo and also use the quantum Fisher
information to quantify the amount of (phase) information lost by
the central spin. We further show that the non-Markovianity measure of
Breuer \textit{et al.} can be used to pinpoint the critical value
of the transverse field.

Let us consider a two-level system (qubit) undergoing a purely
dephasing dynamics due to its coupling to a spin environment.
Assume the initial state of the composite system to be factorized,
$\rho^{tot}(0) =|\phi_{s}\rangle \langle \phi_s|\otimes
\rho^{env}(0)$, with a pure system state
$|\phi_{s}\rangle=c_{g}|g\rangle+c_{e}|e\rangle,\;|c_g|^2+|c_e|^2=1$.
The dynamics of the environment splits into two branches with
weights depending on the state of the qubit. The branches are
characterized by effective Hamiltonians $H_{\alpha} = H_{env} +
\bra{\alpha}H_{int}\ket{\alpha}$, where $\alpha=e,g$ and $H_{env}$
and $H_{int}$ are the environment Hamiltonian and the interaction
Hamiltonian, respectively. The effective Hamiltonian for each
branch includes a perturbation induced by the system in state
$|\alpha\rangle$, which gives rise to a back-action on the
environment. Under the action of this environment the qubit
evolves as $\rho_{s}(t)= |c_{g}|^2|g\rangle\langle g|
+|c_{e}|^{2}|e\rangle\langle e|
+c_{g}^{*}c_{e}\nu(t)|e\rangle\langle g| + \mbox{H.c.}$, where
$\nu(t)$ is the so-called decoherence factor. If the initial
environmental state is pure, $\rho^{env}(0) =
|\Phi\rangle\langle\Phi|$, the decoherence factor is simply the
overlap between perturbed environmental states of the two branches
$\nu(t)= \langle\Phi|e^{i \hat H_{g}t}e^{-i\hat
H_{e}t}|\Phi\rangle$ (we set $\hbar=1$). Finally, the square of
the decoherence factor gives a quantity known as the Loschmidt
echo:
\begin{equation}
L(t)=|\nu(t)|^{2}=|\langle\Phi|e^{i \hat H_{g}t}e^{-i\hat H_{e}t}|\Phi\rangle|^2.
\end{equation}
The Loschmidt echo describes the stiffness of the environment with
respect to the specific system perturbation and it relates to the
decoherence of the qubit in the following simple way: consider the
purity of the qubit $P(t)=\text{Tr}_s(\rho_s^{2})$; for an initial
equatorial spin state this reduces to $P(t)=[1+L(t)]/2$. When
$L(t)\rightarrow 0$ then $P(t)\rightarrow\frac{1}{2}$, indicating
that the qubit is maximally entangled with the environment and
signalling a complete loss of coherence.

In order to explicitly connect the Loschmidt echo to
information flux we
use the approach to the quantification of non-Markovianity put
forward by Breuer {\it et al.}, which is based on the information
flow between a system and its environment \cite{elsi:09}. This, in turn,
is given by the rate of change of the trace distance between
different system states. For a pair of system states $\rho_{1,2}$
the trace distance is
$D(\rho_{1},\rho_{2})=Tr|\rho_{1}-\rho_{2}|/2$. It yields a
natural metric on the state space, invariant under unitary
transformations and not increasing under dynamical (completely positive and
trace-preserving) maps and tells us how distinguishable the two
states are. In a Markovian process the distinguishability between
any two quantum states decreases monotonously, indicating a loss
of information, whereas a non-Markovian process is characterized
by its growth for some time interval in which information flows
back to the system. In terms of the rate of change of the
distinguishability,
$\sigma[t,\rho_{1,2}(0)]=dD[\rho_{1}(t),\rho_{2}(t)]/dt$, where
$\rho_{1,2}(t)$ are time evolved states, a non-Markovian process
must have $\sigma>0$ for some time interval. One can define the
degree of non-Markovianity of a quantum process by optimizing over
all possible pairs of input states:
\begin{equation}
 \label{eq:N}
\mathcal{N}=\max_{\rho_1,\rho_2}\int_{\sigma>0}dt\;\sigma(t,\rho_{1,2}(0)).
\end{equation}
For the purely dephasing process, such as the one considered in
this Letter, the optimization is achieved with equatorial,
antipodal states \cite{manychinese11}. Interestingly, these are
eigenvectors of a spin component in a direction orthogonal to that
of the pointer states. The trace distance of two such states under
the dephasing noise described in the previous section is
$D[\rho^{1}_{s}(t),\rho^{2}_{s}(t)]=\sqrt{L(t)}$. This gives a
neat expression for the measure of non-Markovianity in terms of
the Loschmidt echo:
\begin{equation}
\label{eq:NL} \mathcal{N}=\sum_{n}
\sqrt{L(b_{n})}-\sqrt{L(a_{n})},
\end{equation}
where $[a_{n},b_{n}]$ are the time intervals over which $L'(t)>0$
and $L(a_n)$ and $L(b_n)$ are local minimum and maximum,
respectively, of the Loschmidt echo. We have thus arrived at a
simple relationship between the Loschmidt echo and the
non-Markovianity measure introduced in Ref. \cite{elsi:09}.
Intuitively, a monotonously decaying $L(t)$ is a signature of
Markovian dynamics, while the presence of a time oscillation is a
direct sign (and its amplitude is a quantitative measure) of
information back-flow.

Further insight in the information dynamics comes from the
master equation for the qubit. From the form of $\rho_s(t)$, one
obtains
\begin{eqnarray}
\frac{d \rho_s(t)}{dt} = i \Lambda (t) [\sigma_z,\rho_s(t)] + \gamma(t)
\left[ \sigma_z \rho_s(t) \sigma_z -  \rho_s(t) \right], \label{eq:ME}
\end{eqnarray}
with $\gamma(t)= -L'(t)/[4 L(t)]$ and  $\Lambda(t)=-\phi'(t)/2$,
where we use notation $\nu(t)=|\nu(t)|e^{i\phi(t)}$. The effect of
the environment is twofold: it gives rise to a time dependent Lamb
shift $\Lambda(t)$ and to a time dependent decay rate $\gamma(t)$.
Since the master equation above is exact, it allows to draw a
number of conclusions. It has a Lindblad-like form, but with a
time dependent rate which may take temporarily negative values,
occurring whenever the Loschmidt echo increases with time. When
this happens the dynamical map $\Phi$ describing the system
dynamics $\rho_s(t)=\Phi(t,t_0)[\rho(t_0)]$ is non-divisible, that
is, it cannot be written as a composition of two dynamical maps
$\Phi(t,t_0)\neq\Phi(t,s)\Phi(s,t_0)$. This makes the dynamics of
the qubit non-Markovian also according to a definition proposed
in Ref. \cite{Plenio:10}.

Indeed, the relation between environmental response to the system
perturbation, quantified by the Loschmidt echo, and information flow can be established by
using other non-Markovianity measures. In the proposal put forward
in Ref. \cite{Plenio:10} one considers the decohering qubit to
share a maximally entangled state with an ancilla. By monitoring
the time evolution of an entanglement measure, $E_m(t)$, one can
define a non-Markovianity measure as ${\cal N}_E = \int_{\dot
E_m>0} \, \dot E_m \, dt$. By choosing concurrence $C$ as the
entanglement measure, for a purely dephasing dynamics one finds
that the two non-Markovianity measures exactly coincide:
\begin{equation}
C(t) \equiv \sqrt{L(t)} \; \;   \Rightarrow \; \; {\cal N}_E = \int_{\dot
L
>0} \frac{d\sqrt{L(t)} }{dt} \, dt \equiv  \mathcal{N} \, .
\end{equation}
Yet another possibility to quantify the non-Markovian character of
the time evolution is given by the Fisher information \cite{fishermarkov}.
The idea is the following: assume that a phase gate $U_{\theta} = \ket g \bra g + e^{i \theta} \, \ket e
\bra e$ is applied on the
input state $\ket{\psi_{in}}$ of the qubit at time $t_0$, before it starts interacting with the dephasing
environment. After a time $t>t_0$, one can try to estimate the
superimposed phase $\theta$. The error performed in any kind of
such unbiased estimations is lower bounded by the Cram\'{e}r-Rao
formula in terms of the quantum Fisher information,
$\mbox{Var}(\theta, t) \geq 1/{\cal F}(t)$. The quantum Fisher information is the best possible accuracy achievable
in estimating the parameter $\theta$: it gives a
quantitative description of how much the extractable (phase)
information contained in the qubit has been deteriorated by the
interaction with the environment.

In the absence of environment, the best choice is to select an
eigenstate of $\sigma_x$ as input state. We assume the same
initial qubit state even in the presence of decoherence and obtain
the quantum Fisher information flow in terms of the Loschmidt echo
and the decay rate $\gamma (t)$ as
\begin{equation}
{\cal I}_F(t) \equiv  {\cal F}'(t) =- 4 \gamma(t) \, L(t).
 \end{equation}
 Since  the Loschmidt echo $L(t)$ is always
positive, this expression gives an intuitive relation between the quantum Fisher
information flow and the decay rate: when $\gamma (t) >0$, phase
information flows out of the qubit to become non-local as it is
recorded in the full system+environment state, while for $\gamma(t)
<0$ there is an information back-flow towards the qubit. These results qualitatively agree with the information quantified
in terms of distinguishability by the trace distance. To make the
agreement quantitative, one could take $\tilde{D} =
D(\rho_s^1(t),\rho_s^2(t))^2$ as the distance between evolved
states, and define a new information flow $\tilde{\sigma}(t) =
\frac{d}{dt} \tilde D$. This newly defined flow, once optimized
over the possible initial states of the system, coincides
with the quantum Fisher information flow $\tilde{\sigma}(t) \equiv {\cal I}_F(t)$.

We now specify the discussion above to the case in which the
environment is described by an one dimensional Ising model in
transverse field \cite{Quan},
\begin{equation}
\label{eq:isingtransverse}
H_{env}(\lambda)=-J\sum_{j}{\sigma^{z}_{j}}{\sigma^{z}_{j+1}}+\lambda{\sigma^{x}_{j}},
\end{equation}
where $J$ is the exchange interaction strength and $\lambda$ describes the strength of the transverse field. We assume that the qubit is centrally coupled to the spin chain via the interaction term
\begin{equation}
H_{int}(\delta)=\delta|e\rangle\langle e|\sum_{j}\sigma^{x}_j \, ,
\end{equation}
leading to a state-dependent transverse field strength
$\lambda^*=\lambda+\delta$, where $\delta$ is the strength at
which the qubit couples to the environment. By means of a
Jordan-Wigner transformation, the Hamiltonians $\hat{H}_{\alpha}$
with $\alpha=g,e$, may be diagonalized by suitable sets of fermion
creation and annihilation operators $\hat{c}^{\alpha}_{k}$ such
that
\begin{equation}
\label{eq:fermiquadratic}
H_{\alpha}=\sum^{N}_{k=1}\epsilon^{\alpha}_{k}({c^{\dagger\alpha}_{k}c^{\alpha}_{k}}-\frac{1}{2}).
\end{equation}
Operators $c^{\alpha}_{k}$ are connected to each other by means of
a Bogoliubov transformation $c^{e}_{k}=\cos(\beta_{k})c_{k}^{g}-i\sin(\beta_{k})c_{-k}^{g}$. After some algebra it can be shown that the the Loschmidt echo for the Ising chain is
\begin{equation}
\label{loschmidtIsing}
L(\lambda^{*},t)=\prod_{k>0}[1-\sin^{2}(2\beta_{k})\sin^{2}(\epsilon^{k}_{e}t)]
\end{equation}
where $\beta_{k}$ are the Bogoliubov angles and $\epsilon^{k}_{e}$
are the single quasiparticle excitation energies of the system
with central spin in the state $|e\rangle$, see Ref. \cite{Quan}.

\begin{figure}[tb]
  \includegraphics[width=0.9\linewidth]{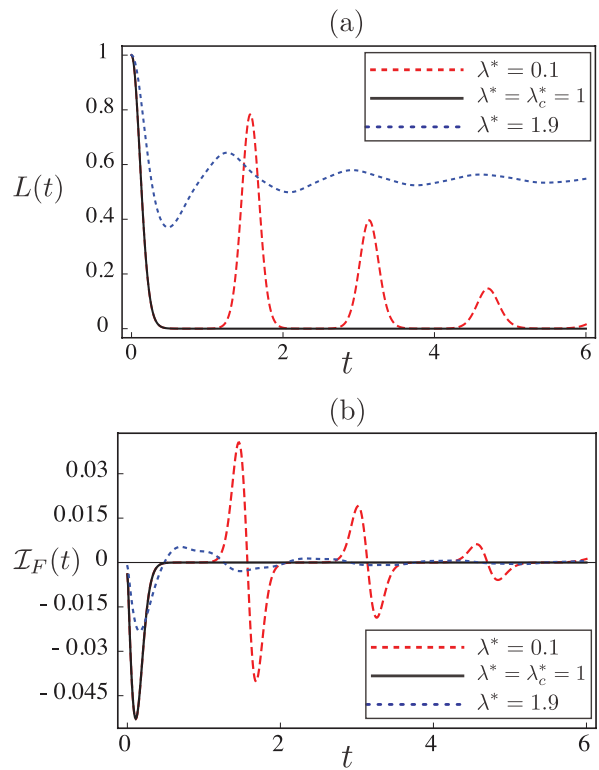}
  \caption{(Color online) (a) The Loschmidt echo and (b) the Fisher information flow induced by the coupling of a central spin to the transverse Ising model consisting of $N=4000$
spins for three different values of the transverse field. We choose $J=1$ and $\delta=0.01$. The sensitivity to the critical point $\lambda^*_c=1$
is dramatic. Note that in Fig. (a) the solid black curve and the dashed red curve overlap for initial times.}
  \label{fig:loschmidtecho}
\end{figure}

 \begin{figure}[tb]
  \includegraphics[width=0.9\linewidth]{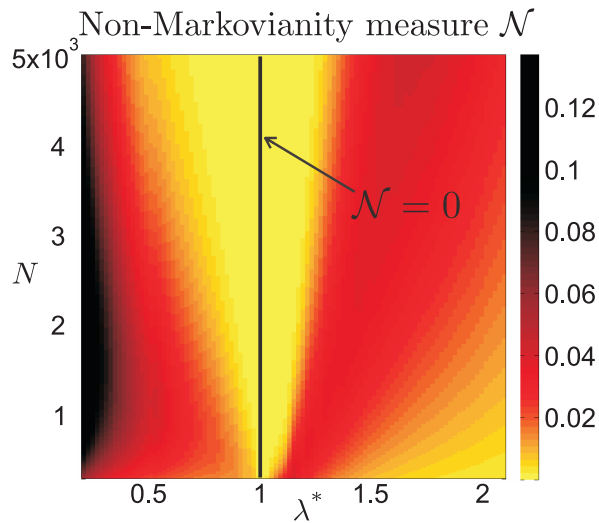}
  \caption{(Color online) Non-Markovianity measure ${\cal N}$ as a
function of the particle number $N$ and the renormalized field
$\lambda^*$. The time-integral in $\mathcal{N}$ has been truncated to avoid the finite-size effects. A black line has been added to highlight the area where the measure is
exactly zero, corresponding to purely Markovian dynamics.}
  \label{fig:nonmarkov}
\end{figure}

For this spin environment the exchange interaction and the
external field tend to make the state rigid in orthogonal
directions. When the two balance the environment is at a critical
point, where it is most susceptible to perturbations. Decoherence of the central qubit
is strongest at this point \cite{Zurek11}, and the decay of
the Loschmidt echo is strongly enhanced \cite{Quan}. We demonstrate this effect in Fig.
1(a) . The Loschmidt echo oscillates strongly outside the critical point, particularly for small values of the transverse field when the exchange interaction dominates the dynamics of the spin chain.

In light of our earlier results, these oscillations correspond to
highly non-Markovian qubit dynamics, characterized by many
intervals of time when the direction of the information flux is
reversed. This is best seen in the flow of quantum Fisher
information, plotted in Fig. 1(b). As pointed
out before, such behavior can be expected when a system is in
contact with a structured, interacting environment such as the
spin chain in a transverse field we consider here. The information
exchange is greatest when the environment is strongly interacting,
that is, the effect of the transverse field is small in comparison
to the spin-spin interaction.

Instead, at the critical value of the transverse field the
Loschmidt echo decays monotonically, pointing to purely Markovian
dynamics of the qubit. The hyper-sensitivity of the Loschmidt echo
to the critical point translates in a striking way to the
non-Markovianity measure, which is shown in Fig.
2. Indeed, the qubit dynamics becomes Markovian
exactly at the critical point, while outside the critical point we
always observe non-Markovian effets.

Notice that ${\cal N}$ is zero for $\lambda^*=1$ even for finite
sized spin environments (in fact, ${\cal N}=0$ for any value of
$N$). Thus, strictly speaking, ${\cal N}$ is
not a critical quantity. However, it gives an indication of the
fact that the environmental fluctuations are slowing down at
criticality, the environment is becoming stiff and is not able to
react on the qubit within a time window smaller than the
recurrence time. This results in a monotonous decay of Loschmidt
echo and, as a consequence of the relation we have found,
Markovian dynamics for the qubit. This slowing down only occurs at
the quantum phase transition point, and therefore, in this sense,
the non-Markovianity measure $\mathcal{N}$ is an indicator of
criticality.

In conclusion, our results provide a unified picture of
decoherence in spin environments by connecting the Loschmidt echo
to the time-dependent dephasing rate of the exact master equation,
and to three different measures of non-Markovianity. This
connection sheds new light on dephasing and einselection in spin
environments, relating these phenomena to non-Markovianity and
memory effects. The dynamics of the Loschmidt echo is shown to be
directly linked to information flux between a qubit and the
environment. Most notably the connection to the Fisher information
reveals how phase information about the qubit is lost, and when it
can be temporarily regained.

We further explored this connection in the context of a qubit
centrally coupled to an Ising spin chain and discovered that the
non-Markovianity measure has a strong imprint of the quantum phase
transition of the Ising model, even for a finite-sized
environment. Indeed  ${\cal N}$ has the remarkable property of
being able to pinpoint the critical value of the transverse field
even away from the thermodynamic limit where the quantum phase
transition truly takes place. This may be very useful in quantum
simulations of the Ising model in finite systems such as trapped
ions, where the number of spins is small \cite{Monroe}.

 \begin{acknowledgments}
P.H., S.Mc., and S.M. acknowledge the Emil Aaltonen Foundation and the Finnish Cultural Foundation (Science Workshop on Entanglement) for financial support.
J.G. would like to acknowledge funding from an IRCSET
Marie Curie International Mobility fellowship and thank T.~Apollaro and M.~Paternostro for interesting discussions.
\end{acknowledgments}

\end{document}